\documentclass[a4paper,12pt]{article}
\usepackage[dvips]{graphicx}
\usepackage{amsmath}
\begin{document}
\begin{center}
{\large \bf On physical vacuum as an unobservable system}
\vspace{5ex}

I. K. Rozgacheva$^a$, A. A. Agapov$^b$
\vspace{2ex}

VINITI RAS, Moscow, Russia
\vspace{1ex}

E-mail: $^a$rozgacheva@yandex.ru, $^b$agapov.87@mail.ru
\vspace{4ex}
\end{center}
\begin{abstract}
Modern astronomical observations in cosmology provide increasingly strong evidence that the expansion of the Universe is accelerating. Explanations of the cosmic acceleration within the framework of general relativity use the hypothesis about a dark energy field (physical vacuum) with unrealistic fine-tuned unnatural properties to the properties of the observable matter. The main difficulty of  the speculative models is that the gravitational properties of the vacuum are unknown. In this work the geometric properties of the physical vacuum which are the consequence of its general property to be unobservable (vacuum does not affect the motions of any real bodies) are considered. It is shown that the effective homogeneous and isotropic space-time of the physical vacuum has four closed dimensions. The vacuum fluctuations create interactions of the real particles. It is shown if we assume the cause principle and there are not any arbitrary real particles births then the lengths of waves of vacuum fluctuations make the fractal manifold. In this case all real systems have the fractal properties.
\end{abstract}
\vspace{1ex}

\vspace{3ex}

\begin{center}
{\bf 1. Introduction}
\end{center}

The concept of the physical vacuum is used for treatment of experimental data in elementary particle physics. The same concept referred to as "dark energy" is used in modern cosmology for treatment observational data on accelerated Universe expansion, motion of galaxies in clusters and galaxy rotation. Assumed properties of the physical vacuum show that it has to be a complex system. The vacuum consist of vacuum fluctuations. From the one hand, the fluctuations are supposed to create interactions of real particles. From the other hand, they are not observable directly, because the general property of the physical vacuum is to be unobservable (it does not affect the motions of any real bodies).

In this work we consider the properties of the vacuum effective geometry in the framework of hypotheses used in the quantum field theory and in the general relativity. The both theories have a convincing experimental foundation. It is shown that the vacuum is a fractal. Therefore, the self-similarity property is expected to characterize real complex systems from atoms to galaxy clusters.
\vspace{2ex}

\begin{center}
{\bf 2. Effective geometry of the physical vacuum}
\end{center}

The state in which there are no real particles and all interaction fields are equal to zero is called the physical vacuum. This concept is very important in the quantum fields theories. It goes back to Poincar\'e's hipothesis of the electromagnetic aether \cite{1}, Einstein's hipothesis of the gravitational aether \cite{2} and Dirac's hipothesis of the electron-positron aether \cite{3}.

The vacuum is supposed to consist of independent quantum fluctuations of interaction fields. They are vacuum fluctuations (fundamental vacuum bosons) for which variation of field energy $\delta E$ during time $\delta t$ satisfies Heisenberg uncertainty relation: $\delta E\cdot\delta t=\hbar$.

Interaction of real particles with vacuum bosons may lead to creation of virtual fermion particle-antiparticle $(q-\bar{q})$ pairs. For this the fluctuation energy must be equal to $\delta E=2mc^2$ (where $m$ is the particle mass) due to interaction. The time of fluctuation energy variation is equal to $\delta t=\hbar/2mc^2$. During this time the virtual $q\bar{q}$-pair disappears. The transformations of vacuum bosons to virtual fermion pairs and vice versa are supposed to happen in the physical vacuum  permanently.

A virtual fermion pair may become a real one if it receives energy during creation faster than energy variation in vacuum fluctuations, i.e.
$$
\Delta t=\frac{\hbar}{E}<\delta t=\frac{\hbar}{2mc^2}.
$$
This hypothesis of particle creation from vacuum bosons is used for explanation of hadron jets observed in accelerators which are produced in electron-positron and proton-antiproton colliding beams. The higher energy of colliding particles the more variety of particles in the jets. It is supposed that colliding particles are surrounded by a “coat” of vacuum bosons. During a collision these bosons transform into flying apart fermion pairs.

The concept of particle creation from vacuum fluctuations allows us to understand the data on radioactive nuclei decays. Decay products are not parts of a radioactive nuclei. They are created during the decay. The concept of vacuum gauge gluons composing a “coat” around a quark is used for explanation of asymptotic freedom of quarks in protons. The concept of vacuum fluctuations of electromagnetic field is used in quantum electrodynamics for explanation of ground energy level shift in isolated hydrogen atom (Lamb shift) \cite{4}, an anomalous magnetic moment of electron, and attraction of two conductive plane-parallel plates in vacuum (Casimir effect) \cite{5}, \cite{6}.

These examples show that the physical vacuum is a cornerstone of modern physics. All real fermions and bosons are considered as excitation states of the vacuum. Interactions of all particles are provided by the vacuum. Consequently, the fundamental physical constants are related to the vacuum properties; particularly, the maximal speed of interaction propagation $c$, the minimal quantum of action $h$, and constants of all kinds of fermion interactions.

Unobservability is the fundamental vacuum property. For the unobservability of the vacuum its dynamical invariants have to be independent on choice of a reference frame on the average. Otherwise, one could reveal the vacuum variation under movement of the reference frame.

Let us consider two dynamical invariants related to space-time properties: energy-momentum 4-vector $P^i$ and angular momentum tensor $M^{ik}$ $(i,k=0, 1, 2, 3)$. It is known that, according to Noether's theorem, in an isolated reference frame the invariance of the 4-vector $P^i$ under four-dimensional space-time shifts is related to the invariance (symmetry) of the action functional under these shifts. The action functional is proportional to the space-time interval, so the symmetry of the action under the shifts is equivalent to the space-time homogeneity. For example, all points of the space-time are equivalent for a point particle moving with a constant moment, and every point may be considered as a reference zero point. Analogously, in an isolated reference frame the invariance of the angular momentum tensor $M^{ik}$ under four-dimensional rotations of the reference frame is related to the space-time isotropy.

We use the geometrical approach and introduce the vacuum space-time for an effective (fluctuation-averaged) description of the vacuum. If the average vacuum values $P^i$ and $M^{ik}$ are equal to zero everywhere the space-time is homogeneous and isotropy, i.e. symmetric under shifts and rotations.

In order no one could reveal relative motion of different vacuum’s parts, simultaneous conservation of $P^i$ and $M^{ik}$ is necessary. In this case, no one can reveal vacuum fluctuation flows under 4-dimensional rotations of reference frame, i.e. there is no nonzero energy-momentum 4-vector $P^i$, and there is no nonzero angular momentum tensor $M^{ik}$ under 4-dimensional shifts.

The vacuum space-time is mapped onto itself under any displacements of the reference frame. In the set topology such mappings are called automorphisms. The automorphisms of the shifts corresponds to the conservation law of the energy-momentum 4-vector $P^i$. The automorphisms of the rotations corresponds to the conservation law of the angular momentum tensor $M^{ik}$. Coincidence of the automorphisms of the vacuum space-time is necessary for its unobservability. This condition increases the vacuum symmetry as compared with the homogeneous and isotropy Minkowski space-time for a free pointlike particle.

The homogeneous and isotropy vacuum space-time has the metric
$$
ds^2=g_{ik}dx^idx^k. \eqno (1)
$$
Its metrical tensor $g_{ik}$ is related to the metrical tensor of the Minkowski’s world $\eta_{ik}$ through the conformal transformation $g_{ik}=a^2\eta_{ik}$, where the diagonal unity matrix $\eta_{ik}$ has the signature $(+ - - -)$ and $a$ is a constant scale factor.

The automorphisms of the manifold $\Phi$ is described through the Killing vector $\xi_i$. During a shift of the manifold $\Phi$ along the $\xi_i$ vector the coordinate transformation $\tilde{x}_i\to x_i+\xi_i$ is performed under the condition that metrical relations between points of the manifold remain constant: $g_{ik}(\tilde{x})=g_{ik}(x)$. Killing vectors tangent space-time $\Phi$ geodesic lines and satisfy the equation \cite{7}
$$
\xi_{i;k}+\xi_{k;i}=0,\eqno (2)
$$
where $\xi_{i;k}$ is a covariant derivation in the metric (1). This equation has a solution for automorphisms of 4-shifts:
$$
\xi_i=T_i,\eqno (3)
$$
where $T_i$ is a constant 4-vector, and for four-dimensional rotations:
$$
\xi_i=V_{ik}x^k,\eqno (4)
$$
where $V_{ik}$ is a constant antisymmetric matrix of four-dimensional rotations.

Let us define the 4-vector $T_i$ as a 4-dimensional gradient at the $\Phi$ surface:
$$
T^i=\frac{\partial\Phi}{\partial x_i}.
$$
The covariant components are $T_i=g_{ik}T^k$. The condition of the automorphisms (3) and (4) coincidence gives the equation for the $\Phi$ function:
$$
\eta_{ik}\frac{\partial\Phi}{\partial x_i}\frac{\partial\Phi}{\partial x_k}=\eta_{ik}V^{il}V^{km}x_lx_m.\eqno (5)
$$

Using the equation (5) we can ascertain that the $\Phi$ manifold is compact. For simplicity we consider two-dimensional automorphisms in the coordinate plane $\{x_0=ct, x_1\}$ (Lorentz rotations). In this case, the indices in the equation (5) range over values 0, 1 and the equation is transformed into the form
$$
\left(\frac{\partial\Phi}{\partial x_0}\right)^2-\left(\frac{\partial\Phi}{\partial x_1}\right)^2=\Psi^2,\eqno (6)
$$
where $\Psi^2=-(aV^{10})^2(x_0x_0-x_1x_1)$. For the Lorentz rotations
$$
V^{10}=i\frac{v/c}{\sqrt{1-\left(v/c\right)^2}},
$$
where $v$ is a velocity of the reference frame movement, $i$ is the imaginary unit. Note that $\Psi^2>0$ for time-like movements $\left(x_0>x_1\right)$. In the equation (6) it is taken into account that due to the antisymmetry the equalities $V^{00}=V^{11}=0$ and $V^{10}=-V^{01}$ are satisfied. We consider that in the reference frame $\{x^0, x^1\}$ the interval is equal to
$$
s^2=g_{ik}x^ix^k=a^2(x^0x^0-x^1x^1),
$$
and
$$
ds^2=g_{ik}dx^idx^k=a^2(dx^0dx^0-dx^1dx^1). \eqno (7)
$$
One can ascertain that equalities (6) and (7) become identical for the functions
$$
\frac{\partial\Phi}{\partial x_0}=\Psi\cosh{\Phi},\ \ \ \frac{\partial\Phi}{\partial x_1}=\Psi\sinh{\Phi},\ \ \ \frac{dx^0}{ds}=\frac{1}{a\Psi}\frac{\partial\Phi}{\partial x_0},\ \ \ \frac{dx^1}{ds}=\frac{1}{a\Psi}\frac{\partial\Phi}{\partial x_1}. \eqno (8)
$$
Equations (8) allow us to find the following relation between $\Phi$ and the interval $s$:
$$
-\frac{i}{2}V^{10}\frac{s^2}{a^2}=\arctan{\left(\mbox{e}^{2\Phi}\right)}-\frac{\pi}{4},\eqno (9)
$$
where the condition $s\left(\Phi=0\right)=0$ is used. The equality (9) shows that for an infinite manifold of points $0\le\Phi\le\infty$ the interval values range over finite volume
$$
0\le s\le \left(-\frac{\pi}{2iV^{10}}\right)^{1/2}a=s_{max}.\eqno (10)
$$

Using equations (8) and solution (9) one can derive $x_0(s)$ and $x_1(s)$ functions. It is found that variations of the coordinates $x_0$ and $x_1$ are also finite. In order to coordinate lines have no boundaries they have to be closed. Therefore, a coordinate network on a two-dimensional manifold $\Phi$ is described by two orthogonal circles $C(x_0)$ and $C(x_1)$. Each circle is defined through identifying of two boundary values of the corresponding coordinates $x_0$ and $x_1$. In this case, the $\Phi$ manifold is a torus, where the coordinate circle $C(x_0)$ is a parallel of the torus and $C(x_1)$ is its meridian.

It is important that closing the coordinate lines $x_0(s)$ and $x_1(s)$ we identify the world points $s=0$ and $s=s_{max}$. That means that the $\Phi$ manifold contains closed geodesic lines and it is compact, although it consists of infinite numbers of geometric points.

Analogously, one can ascertain using equation (5) that the surfaces $\{x_0, x_2\}$, $\{x_0, x_3\}$, $\{x_1, x_2\}$, $\{x_1, x_3\}$, $\{x_2, x_3\}$ are also compact.
\vspace{2ex}

\begin{center}
{\bf 3. Stochastic property of physical vacuum}
\end{center}

Here we make sure of stochastic property of the vacuum geometry. The physical concept of stochasticity of dynamical systems is for instability. In terms of geometry the concept is considered as the instability of phase trajectory of system under the external influence. The treatment was appeared in Kovalevskaya’s paper \cite{8}.

The effective phase trajectories of the vacuum fluctuations are the geodesic lines of the $\Phi$ manifold. The instability of the geodesic lines is explored by means of the Jacobi equation.

Let us consider two-dimensional surface $\Phi(x_0, x_1)$ and the geodesic variation $\sigma(s)$ with the interval $s$ along the geodesic line. The Jacobi equation may be written as \cite{9}:
$$
\frac{d^2\sigma}{ds^2}=-K\sigma,\eqno (11)
$$
where $K$ is the Riemannian curvature of the surface $\Phi(x_0, x_1)$ in the world point $s$. The geodesic variation $\sigma$ defines the deviation of geodesic line from its initial direction.

The curvature $K$ of the surface $\Phi(x_0, x_1)$ is calculated from the formula:
$$
K=\frac{\displaystyle\frac{\partial^2\Phi}{\partial x_0^2}\frac{\partial^2\Phi}{\partial x_1^2}-\left(\frac{\partial^2\Phi}{\partial x_0\partial x_1}\right)^2}{\displaystyle\left[1+\left(\frac{\partial\Phi}{\partial x_0}\right)^2+\left(\frac{\partial\Phi}{\partial x_1}\right)^2\right]^2}.\eqno (12)
$$

If the curvature $K$ is independent of the interval $s$ and $K=K_0<0$ then from the equation (11) we find the exponential solution ${\displaystyle \sigma\propto\mbox{e}^{\sqrt{-K_0}s}}$.

Using the formulae (8), (9), (12) we can obtain for $K<0$ and the Lorentz's transformations:
$$
K\approx-\frac{(v/c)^2}{1-(v/c)^2}\cdot\frac{s}{a^3}.\eqno (13)
$$

Hence the curvature of the compact surface $\Phi$ is negative, and the deviation of geodesic lines increase following the low $\sigma\propto\sinh{(ks)}$, where
$$
k\approx\left(\frac{v/c}{\sqrt{1-(v/c)^2}}\right)^{2/3}\frac1a
$$
and $ks\ll1$. Such result one can obtain for the surfaces $\Phi(x_0, x_2)$, $\Phi(x_0, x_3)$.

The deviation of geodesic lines at the compact surface $\Phi$ is the sufficiency condition of mixing of phase trajectories \cite{9}. The mixing leads to the fast loss of information on the initial directions of geodesic lines. Such system is stochastic.

The stochastic geometry makes it possible to reconcile the compactness of the space-time with the causality principle. We cannot leave a message in the past because of the instability of the closed geodesic lines.

Note that with the vacuum fluctuation momentum
$$
p_i=\frac{mc}{\Psi}\xi_i=\frac{mc}{\Psi}\frac{\partial\Phi}{\partial x_i}   \eqno(14)
$$
the equation (6) becomes the mass surface equation:
$$
p_0p_0-p_1p_1=(mc)^2,\eqno (15)
$$
where $m$ is the mass parameter.

As known, the mass surface equation (15) is correct for free point particle which has conserved energy and momentum. Thus from the phenomenological viewpoint the vacuum is system of noninteracting particles with momenta (14). The particles trajectories mixed in a time $\tau=a/kc$.

We must stress the mass surface equation (15) is result from the unobservability of the physical vacuum. Whereas in contrast in the special theory of relativity this equation is consequence of the invariance of interval (1) with respect to coordinate transformations.

Let us consider the connection between the closed space-time of the physical vacuum with the instable geodesic lines and the postulates of quantum mechanics, i.e. stochastic nature of matter and quantized energy of an isolated system.

The mixing scale of the instable vacuum geodesic lines is
$$
\lambda_*=\frac{a}{k}=a\left(\frac{1-(v/c)^2}{(v/c)^2}\right)^{1/3}.\eqno (16)
$$
The behavior of virtual particles at scales $\lambda>\lambda_*$ is random nature. At scales $\lambda<\lambda_*$ vacuum the vacuum geodesic lines are stable, i.e. there are no vacuum fluctuations.

Let us suppose the scale (15) is an analog of the Broglie wave length $\lambda_*\to\hbar/mv$. Using the Compton wave length $\lambda_0=\hbar/mc<\lambda_*$ we can obtain from (16) an algebraic equation for value $\lambda_0/\lambda_*=v/c=y$:
$$
y^3-y+\left(\frac{\lambda_0}{a}\right)^3=0.\eqno (17)
$$

Next we take into account the closedness of geodesic lines. In order that there are no the spontaneous birth of particles from vacuum it is necessary an integer number $N$ of the Broglie wave length $\lambda_*$ at geodesic length $s_{max}$:
$$
s_{max}=N\lambda_*.\eqno (18)
$$
From conditions (10) and (18) and equation (16) one can obtain for $N\gg1$ the equation:
$$
\lambda_0\propto\frac{s_{max}}{N}. \eqno (19)
$$

As is seen from formula (19) the range of length $\lambda_0$ is discrete. Thus the range of vacuum fluctuation energies $E=mc^2=\hbar c/\lambda_0$ is the same discrete.
\vspace{2ex}

\begin{center}
{\bf 4. Conclusion}
\end{center}

This work has shown the stochastic nature of physical vacuum follow from its unobservability. The interactions of real complex systems are caused by vacuum fluctuations with quantized energy. It is possible, just for this reason the systems with fractal properties are common in the Universe \cite{10}, \cite{11}, \cite{12}, \cite{13}, \cite{14}, \cite{15}, \cite{16}. The fractal cosmological model with the vacuum as the Goldstone boson was described in our paper \cite{17}.

\end{document}